\documentclass[article]{aa}  
\usepackage{graphicx}  
\usepackage{txfonts}  
\bibpunct{(}{)}{;}{a}{}{,} 

\begin{document}  

\title{Chromospheric activity and evolutionary age of   
the Sun and four solar twins}  
  
\author{M. Mittag\inst{1} \and K.-P. Schr\"oder\inst{2} \and A.   
Hempelmann\inst{1} \and J.N. Gonz\'alez-P\'erez\inst{1} \and J.H.M.M.   
Schmitt\inst{1}}  
  
\institute{Hamburger Sternwarte, Universit\"at Hamburg, Gojenbergsweg 112,   
21029 Hamburg, Germany\\  
           \email{mmittag@hs.uni-hamburg.de}  
           \and   
           Departamento de Astronomia, Universidad de Guanajuato, Mexico}  
  
\date{Received 12 October 2015; accepted Accepted 8 March 2016}  
  
\abstract  
{}   
{The activity levels of the solar-twin candidates HD~101364 and HD~197027   
are measured and compared with the Sun, the known solar twin 18~Sco, and the   
solar-like star 51~Peg. Furthermore, the absolute ages of these   
five objects are estimated from their positions in the HR diagram and    
the evolutionary (relative) age compared with their activity levels.}  
{To represent the activity level of these stars, the Mount   
Wilson S-indices were used. To obtain consistent ages and evolutionary   
advance on the main sequence, we used evolutionary tracks calculated with   
the Cambridge Stellar Evolution Code.}  
{From our spectroscopic observations of HD~101364 and HD~197027 and based on   
the established calibration procedures, the respective Mount Wilson   
S-indices   
are determined. We find that the chromospheric activity of both stars is   
comparable with the present activity level of the Sun and that of 18~Sco,   
at least for the period in consideration. Furthermore, the absolute   
age of HD~101364, HD~197027, 51~Peg, and 18~Sco are found to be 7.2, 7.1,   
6.1, and 5.1 Gyr, respectively.}  
{With the exception of 51~Peg, which has a significantly higher metallicity   
and a mass higher by about 10\% than the Sun, the present Sun and its twins compare   
relatively well in their activity levels, even though the other twins are somewhat older. 
Even though 51~Peg has a similar age of 6.1 Gyr, this star is significantly less active. 
Only when we compare it on a relative age scale (which is about 20\% shorter for 51~Peg than
for the Sun in absolute terms)   
and use the higher-than-present long-term S$_{\rm{MWO}}$ average of 0.18 for the Sun, 
does the S-index show a good correlation with evolutionary (relative) age. This shows that 
in the search for a suitably similar solar twin, the  
relative main-sequence age matters for obtaining a comparable activity level.}  
  
\titlerunning{Chromospheric activity and evolutionary age for   
the Sun and four solar twins}  
  
\maketitle  
  
\section{Introduction}    
Solar twins are very important since they  
allow astrophysicists to look into the past and future of the  
Sun.
For a star to be considered as a solar twin, its respective stellar   
parameters must be very similar to those of the Sun.   
It is not sufficient that the candidate star be located in the same   
place of the empirical HR diagram, the candidate star must also have a metallicity   
similar to that of the Sun and possess a similar evolutionary age.   
For young stars, sensitive age indicators are rotation, magnetic   
activity \citep[e.g.][]{Mamajek2008ApJ687.1264M},
and lithium abundance, all of which considerably diminish during   
the main-sequence (MS)   
lifetime of a star. By contrast, evolutionary effects become more   
visible in the HR diagram only in the second half of this phase, just   
when the former indicators become less and less sensitive and more ambiguous.    
  
One of the best solar twins appears to be 18~Sco, a star quite   
similar to our Sun   
\citep{Porto_de_Mello1997ApJ...482L..89P}.  18~Sco was used as   
comparison star   
in a study of HD~101364 by \cite{Melendez2012A&A..543A.29M}, who described  
HD~101364 as a ``remarkable solar twin'' with an age estimate   
of (3.5$\pm$0.7) Gyr from the lithium abundance, not too far from the age of  
18~Sco, which \cite{Melendez2012A&A..543A.29M}   
estimated as 2.7 Gyr.  
\cite{Monroe2013ApJ774L32M} presented the star  
HD~197027 as a solar twin with an age of 8.2 Gyr, while, by comparison,   
our Sun only has an age of $\approx$4.5 Gyr, which is a factor
$\text{of about }$2 lower   
than the above age estimate of HD~197027; the same authors   
estimated an age of 2.9 Gyr for 18~Sco.   
The same star was studied by \cite{Ramirez2014A&A572A48R}, who   
presented log~R$^{'}_{\rm{HK}}$ values and ages   
for 18~Sco and HD~197027. While their log~R$^{'}_{\rm{HK}}$   
values are similar, the ages are not, 
since they estimated the age of 18~Sco to be 3 Gyr and that of HD~197027 to be 6.7 Gyr.
  
To study the later activity evolution of our Sun, a   
more evolved MS star is clearly desirable.   
We therefore selected 51~Peg,   
a star with a somewhat higher mass and metallicity than the Sun. This higher mass and metallicity  
mean that 51 Peg evolved somewhat faster   
than the Sun. In this way, 51~Peg serves as a good reference for the future   
solar activity evolution as a late MS star even though
its  
absolute age is comparable.  
  
To study the activity--age relation, a reliable mean S$_{\rm{MWO}}$-index is needed 
to represent the mean level of chromospheric activity of the star.   
Therefore, measurements over a long   
period are required because chromospheric activity is a phenomenon exhibiting  
variability on both short and longer timescales. The Sun, the solar   
twin 18~Sco, and the evolved solar-like star 51~Peg are all included in our   
stellar activity long-term monitoring program to study chromospheric   
activity with our TIGRE telescope. We therefore have obtained 
quite a large number of S$_{\rm{MWO}}$ values for these three objects. 

To compare the chromospheric activity of the Sun, the solar   
twin 18~Sco, the evolved solar-like star 51~Peg, HD~101364, and HD~197027, we   
can directly use the S-index because all these stars and the Sun  
have very comparable $B-V$ colour indices.   
In this case, the Mount Wilson   
S-index is suited best because it is the more empirical and less   
$B-V$-dependent activity indicator compared to the log~R$^{'}_{\rm{HK}}$ value.    
In addition, we estimate the ages of all four object by matching their HR diagram positions   
with suitable evolutionary tracks to cross check these results with the determined activity levels.   
Interestingly, recent studies \citep{Reiners2012ApJ...746...43R,schroeder2013A&A554A50S}  
of solar-like MS stars with some range   
of mass showed that the relative age of a star, that is, its absolute 
age $\tau$ with respect to some reference age,  compares best with the activity level  
and not with the absolute age itself; we return to this point
below.

\begin{figure*}  
\centering  
\includegraphics[scale=0.4]{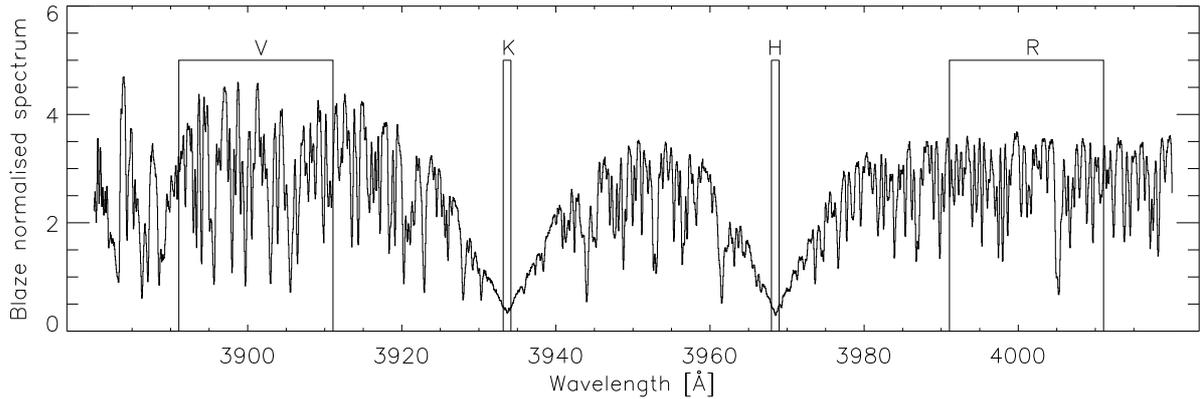}  
\caption{TIGRE spectrum of the Sun (moonlight) with the band-passes   
used in the S$_{\rm{TIGRE}}$ measurement.}  
\label{Fig1}  
\end{figure*}

\section{Observation and data reduction}  
  
During the past years, we have regularly observed the Sun   
(to be precise, we took moonlight spectra), 18~Sco,   
51~Peg, HD~101364, and HD~197027 with the TIGRE telescope, which
is a 1.2~m robotic   
telescope located at the La Luz Observatory on a plateau in central   
Mexico. TIGRE's only instrument is its fibre-fed \'Echelle spectrograph   
HEROS, which allows for simultaneous exposures in two spectral channels   
(a blue and red spectral range) with a   
total spectral coverage from $\approx$3800~\AA~to 8800~\AA~with   
a gap of $\approx$100~\AA~ centred on 5800~\AA~ and a spectral resolution   
of R$\approx$20,000; a more detailed description of the TIGRE facility  
is given by \cite{Schmitt2014AN335787S}.  
  
The observational epochs, the number of observations, the visual magnitude,   
\begin{table}[!t]  
\caption{Object and observational information.}  
\label{tab1}  
\begin{center}  
\begin{small}  
\begin{tabular}{lccccc}  
\hline  
\hline  
\noalign{\smallskip}  
Name & Epoch & No. & B-V & vmag &  $\langle$S/N$\rangle$ \\  
\hline  
\noalign{\smallskip}  
Sun (Moon)  & 08/13 - 05/15 & 184 & 0.65 &  --  & 71 \\  
18~Sco     & 01/14 - 05/15 &  58 & 0.65 & 5.49 & 48  \\  
51~Peg     & 08/13 - 11/14 &  25 & 0.67 & 5.45 & 69  \\  
HD~101364  & 02/15         &   6 & 0.65 & 8.67 & 41 \\  
HD~197027  & 08/14 - 10/14 &   6 & 0.65 & 9.18 & 35 \\  
\hline  
\end{tabular}
\tablefoot{The table includes a list of epochs, number of available spectra, the $B-V$,  
the visual magnitude, and  the mean signal-to-noise ratios (S/N) (in the Ca~II~H\&K 
region (3880-4020 \AA)). The $B-V$ values and visual magnitudes   
are taken from \cite{HIPPARCOS1997ESA}, except for the Sun. The solar  
$B-V$-value is taken from \cite{Cox2000asqu}.}
\end{small}  
\end{center}  
\end{table}  
and the mean S/N in the Ca II H\&K region (3880-4020 \AA) are provided in   
Table \ref{tab1} for the stars we studied here.  
All spectroscopic data were reduced with the TIGRE/HEROS standard reduction   
pipeline, which is implemented in IDL and based on the IDL reduction package   
REDUCE written by \cite{REDUCE2002A&A385.1095P}. The TIGRE/HEROS standard   
reduction pipeline is a fully automatic pipeline with an automatic   
wavelength calibration based on ThAr spectra. This pipeline includes   
all necessary reduction steps to reduce \'Echelle spectra (bias subtraction,   
order definition, wavelength calibration, spectral extraction, and flat fielding).   
  
Dark current and its noise are very low in our data. Therefore, the dark   
subtraction was neglected. We removed the mean dark level by a precise spectrum background correction. The dark noise is   
generally lower than the readout noise and thus negligible as well.   
A master flat-field image created as the average image from   
individual exposures of a tungsten lamp taken at the beginning and the  
end of an observing night.  
For flat fielding the spectrum from the master flat-field image (thereafter: blaze) 
was extracted and the science spectrum was   
divided by the corresponding blaze. With this step, the blaze function is   
removed and the correction of the quantum efficiency is achieved. 
The fringes in the spectrum of the red channel are also removed.   
To obtain the final spectrum that covers a wide range of wavelengths,  
all single extracted and blaze-normalised spectral orders were combined and the  
overlap regions were merged.   
  
\section{Mount Wilson S-index calibration}\label{s_cal}  
  
The so-called Mount Wilson S-index (hereafter S$_{\rm{MWO}}$)   
is a very important and widely used activity indicator for cool stars.   
For a large number of stars and the Sun, S-index measurements have been available  
since the 1960s, in some cases covering up to five decades of monitoring  
undertaken primarily at Mt. Wilson, but also at the Lick and Lowell observatories.  
  
The S-index is based on the chromospheric emission in the Ca~II~H\&K line   
cores, which has originally  
been defined as the ratio of the counts in a triangular band-pass with a FWHM of   
1.09 \AA~ centred on the Ca~II~H\&K lines, denoted by $N_{H}$ and $N_{K}$   
in the following, over the counts   
of two 20~\AA~wide band-passes centred on 3901.07~\AA~and 4001.07~\AA,   
denoted by $N_{R}$ and $N_{V}$, respectively. A multiplicative factor   
$\alpha$ \citep{Vaughan1978PASP90267V,duncan1991} is introduced   
to adjust different instruments to the same S-index scale, so that  
the definition of the S$_{\rm{MWO}}$ is \citep{Vaughan1978PASP90267V}  
  
\begin{eqnarray}\label{sindex_mwo}  
S_{\rm{MWO}} = \alpha \left( \frac{N_{H}+N_{K}}{N_{R}+N_{V}} \right).  
\end{eqnarray}     
  
One of the key projects pursued by the TIGRE telescope is monitoring the   
stellar activity of a sizable stellar sample in Ca~II~H\&K and  
other chromospheric lines.   
To measure the activity, we first determined our own TIGRE S-index, which we   
call  S$_{\rm{TIGRE}}$ in the following,  
from the Ca~II~H\&K lines. These S$_{\rm{TIGRE}}$ values were   
calculated from the blaze-normalised and echelle-order-merged spectra,  
after first performing a barycentric velocity and a radial velocity   
correction of the Ca~II~H\&K region. The barycentric velocity shift   
was estimated as part of the data reduction process and stored in the header   
of each spectrum.   
To obtain the radial velocity shift of the Ca~II~H\&K region, each spectrum   
was compared, guided by a cross correlation,   
with a synthetic spectrum calculated with the atmospheric model code   
PHOENIX \citep{hauschildt1999}. For a realistic application of these   
synthetic spectra, the rotational line-broadening of each object by   
v$\cdot$sin(i)  
and the resolution of the spectrograph (R$\approx$20,000) were   
taken into account. After the radial velocity correction, the counts   
in all four band passes were integrated and the ratio   
between the counts of the line centres and those from the  
quasi-continua were calculated. The error of the S$_{\rm{TIGRE}}$-index   
was determined by error propagation of the standard deviation of the   
four integrals. These integral variations were estimated by means of a   
Monte Carlo simulation, where the S/N of a single   
pixel was considered.   
As an example of a typical TIGRE spectrum in the Ca~II~H\&K region, we show   
in Fig. \ref{Fig1} a spectrum of the Sun (i.e., of moonlight);  
the band-passes defining the S-index are included for illustration.   
  
A main difference between the S$_{\rm{TIGRE}}$ index and the   
classical S-index is that the TIGRE S-measurements are performed with a   
rectangular (not triangular as for Mt. Wilson) band-pass of 1~\AA~at the   
line centre. To compare our numerically integrated results with   
the original Mt. Wilson measurements, we must transform the   
S$_{\rm{TIGRE}}$ values to the Mt. Wilson system by considering
a set of calibration stars (essentially to determine the $\alpha$   
parameter in Eq. \ref{sindex_mwo}).
  
For this purpose, we regularly observed a sample of 50 stars with very well   
\begin{table}[!t]  
\caption{Set of calibration stars observed regularly to guarantee   
the correct transformation of the TIGRE S-index into the Mount Wilson   
S-index.}  
\label{tab2}  
\smallskip  
\begin{center}  
\begin{tabular}{lccl}  
\hline \hline  
\noalign{\smallskip}  
HD~6920$^1$    &    HD~32923$^2$       &    HD~89744$^1$      &   HD~131156A$^1$   \\  
HD~10307$^2$   &    HD~35296$^2$       &    HD~95735$^1$      &   HD~137107$^1$    \\  
HD~10700$^1$   &    HD~37394$^1$       &    HD~97334$^1$      &   HD~142373$^1$    \\  
HD~13421$^1$   &    HD~39587$^2$       &    HD~100563$^1$     &   HD~158614$^1$    \\  
HD~16673$^1$   &    HD~41330$^2$       &    HD~101501$^1$     &   HD~159332$^1$    \\  
HD~17925$^1$   &    HD~42807$^2$       &    HD~106516$^1$     &   HD~182572$^1$    \\  
HD~19373$^2$   &    HD~43587$^2$       &    HD~114378$^1$     &   HD~190360$^1$    \\  
HD~22049$^1$   &    HD~45067$^1$       &    HD~115043$^1$     &   HD~201091$^0$    \\  
HD~22072$^1$   &    HD~61421$^1$       &    HD~115383$^1$     &   HD~207978$^1$    \\  
HD~23249$^1$   &    HD~72905$^1$       &    HD~115617$^1$     &   HD~216385$^1$    \\  
HD~25998$^1$   &    HD~75332$^1$       &    HD~124570$^1$     &   HD~217014$^1$    \\  
HD~26923$^1$   &    HD~75528$^2$       &    HD~124850$^1$     &                   \\  
HD~30495$^1$   &    HD~82443$^1$       &    HD~129333$^1$     &                   \\  
\noalign{\smallskip}  
\hline  
\end{tabular}  
\tablebib{(0)~Extrapolated from time series, corresponding period   
from \citet{b95}, (1)~\citet{b95}, and (2)~Hall (private communication).  
}  
\end{center}  
\end{table}  
known S$_{\rm{MWO}}$ values. These stars (see  Table \ref{tab2})   
were selected from the lists of \citet{b95} and   
\citet{hall2007AJ133862H}, as they are known to be relatively   
constant in their level of activity, that is, do not show any clear or  
significant periodic behaviour in their chromospheric   
emission. A borderline case with a period, however, is the  
very active star HD~201091 (see the Mount Wilson time series from   
\citet[Fig. 1g for HD~201091]{b95}). As the reference Mount Wilson   
S-value for any required epoch, we here used a simple sine function   
with a period of 7.3 yr, an amplitude in S$_{\rm{MWO}}$ of 0.05, and a   
baseline value of 0.658.  
  
\begin{figure}  
\centering  
\includegraphics[scale=0.45]{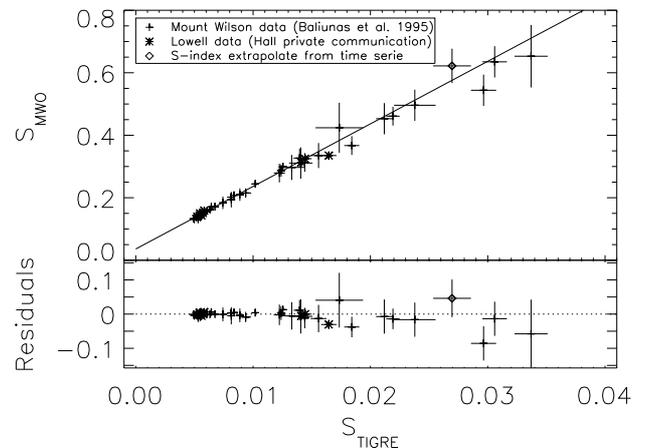}  
\caption{S$_{\rm{MWO}}$ index vs. S$_{\rm{TIGRE}}$ index. Upper panel:   
S$_{\rm{MWO}}$ index vs. S$_{\rm{TIGRE}}$ index, the solid line shows the   
best linear fit, which converts the S$_{\rm{TIGRE}}$ index into   
the calibrated S$_{\rm{MWO}}$ index. Lower panel: the residuals.}  
\label{Fig2}  
\end{figure}

The TIGRE spectra for the objects listed in Table \ref{tab2} 
were obtained between August 2013 and May 2015.   
To determine reliable mean S$_{\rm{TIGRE}}$ values for this period, each star was   
observed at least ten times. In Fig. \ref{Fig2} we compare our results to   
the literature S$_{\rm{MWO}}$ values by plotting S$_{\rm{MWO}}$ values vs. our   
S$_{\rm{TIGRE}}$ values.  
A clear linear correlation is obvious, with a regression curve of    
\begin{eqnarray} \label{trans_eq_hrt_s_mw_index}   
S_{\rm{MWO}} & = & (0.0360\pm0.0029)+(20.02\pm0.42) S_{\rm{TIGRE}}.  
\end{eqnarray}  
The solid line in Fig. \ref{Fig2} represents this regression curve,   
with a standard deviation of the residuals of 0.02. Furthermore,   
the percentage variation between the calculated   
and the corresponding reference   
S$_{\rm{MWO}}$ is only 3.3$\%$.  
  
From Fig. \ref{Fig2} it is obvious that the scatter is smaller for the   
\begin{figure}  
\centering  
\includegraphics[scale=0.5]{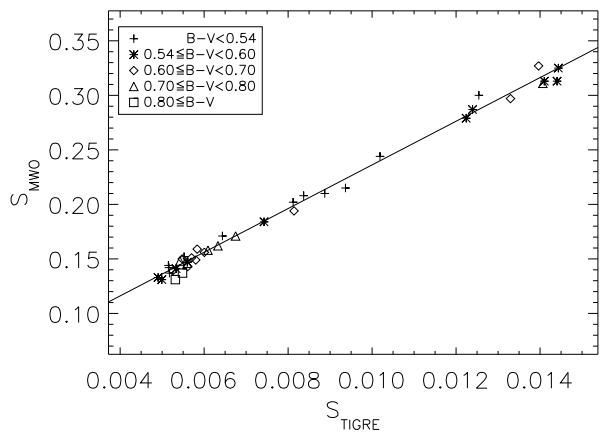}  
\caption{S$_{\rm{MWO}}$-index vs. S$_{\rm{TIGRE}}$-index. Zoom into the relation, and the   
S values are labelled for the different $B - V$ ranges. The solid line shows the linear relation.}  
\label{Fig3}  
\end{figure}  
lower S$_{\rm{MWO}}$ values (S$_{\rm{TIGRE}} \la 0.015$ or S$_{\rm{MWO}} \la 0.36$)   
than for the more active cases. We estimate a scatter between the calculated   
and the original reference values of S$_{\rm{MWO}}$ of only 2.3$\%$ for   
S$_{\rm{TIGRE}}<0.015$, but 6.8$\%$ (a scatter three times larger) for the   
values above.   
Fortunately, the here discussed solar twin stars are all in   
the lower range  (S$_{\rm{MWO}} \la 0.2$) and so are not affected. The   
reason for this less certain  
transformation of S$_{\rm{TIGRE}}$ to S$_{\rm{MWO}}$ values in the active   
range must obviously be attributed to the fact that more active stars,   
even the non-periodic ones,   
are significantly less stable over long timescales than the stars of a   
very modest activity level, which emit chromospheric fluxes much closer to the   
basal level.  
However, with longer time series we will eventually also increase the accuracy   
of the transformation for these more active stars.  
  
Additionally, the S$_{\rm{TIGRE}}$ indices of the calibration   
stars were checked for the colour dependence in the relation between the   
indices S$_{\rm{TIGRE}}$ and S$_{\rm{MWO}}$. Therefore, only low-activity   
stars with S$_{\rm{TIGRE}}<0.015$ were used. These stars were divided into   
different $B - V$ ranges and are plotted in Fig. \ref{Fig3}, where the stars   
are labelled according to the different $B - V$ ranges;   
we find no significant colour dependence of the   
relation between the S$_{\rm{TIGRE}}$ and S$_{\rm{MWO}}$ indices. 
  
Finally, the regression line between the S$_{\rm{TIGRE}}$   
and S$_{\rm{MWO}}$ indices does not pass exactly through the origin, contrary to  
what might have been expected. The reason for this may be the slightly   
different determination of the S values. 
Keep in mind that the lowest real S-values come from stars where the S-values
are at the basal flux level. For giants, the basal flux level is S$_{\rm{MWO}}$=0.086, which is   
higher by than a factor 2 than the constant value of 0.036. For the MS stars these differences 
between basal flux and the constant are even larger \citep{mittag2013A&A549A117M}. 
This constant offset in the calibrated transformation   
relation is not physically relevant and is much lower than the basal   
flux level.
  
\section{Chromospheric activity}  
  
To compare the activity levels of HD~101364 and HD~197027 with the known   
solar-like stars 18~Sco and 51~Peg and with the Sun itself, we used  
their calibrated Mount Wilson S-index values. The S$_{\rm{TIGRE}}$ values   
for the Sun, 18~Sco, 51~Peg, and HD~101364 were determined automatically   
by a tool in the standard TIGRE data reduction pipeline, according to the   
description given in Sect.~\ref{s_cal}, but for HD~197027  
an additional outlier correction was necessary before   
the S$_{\rm{TIGRE}}$ could be measured with the tool described in  
Sect.~\ref{s_cal}.   
The reason for this additional correction is the relatively low S/N of the   
spectra in comparison to the other sample objects.   
Because HD~197027 is the faintest star in our  
sample, it suffers most when observing time is lost due to unstable weather conditions.  
This change was also taken into account for averaging the   
S$_{\rm{TIGRE}}$ indices for HD~197027. Here, the single   
S$_{\rm{TIGRE}}$-index values were weighted for averaging them
with the   
corresponding S/N in the Ca~II~H\&K region. 
  
The mean S$_{\rm{TIGRE}}$ values for the five objects were   
transformed into the Mount Wilson S-scale using  
Eq.~\ref{trans_eq_hrt_s_mw_index}, and the results are listed in   
Table~\ref{tab3}.   
\begin{table}[!t]  
\caption{Results of the the S-index measurements with TIGRE. We list   
the mean S$_{\rm{MWO}}$, the number of spectra, the standard   
deviation ($\sigma$), and the mean literature S$_{\rm{MWO}}$-values.}  
\label{tab3}  
\smallskip  
\begin{center}  
\begin{tabular}{lcccl}  
\hline  
\hline  
\noalign{\smallskip}  
Name &  $\langle$S$_{\rm{MWO}}\rangle$ & No. & $\sigma$ &  Lit. $\langle$S$_{\rm{MWO}}\rangle$ \\  
\hline  
\noalign{\smallskip}  
Sun & 0.172$\pm$0.001 & 184 & 0.003 & 0.179$^{1}$\\  
18~Sco & 0.169$\pm$0.001 & 59 & 0.005 & 0.171$^{2,3,4,5,6,7,9}$\\  
51~Peg & 0.152$\pm$0.001 & 25 & 0.002 & 0.149$^{1}$\\  
HD~101364 & 0.176$\pm$0.003 & 6 & 0.007 & 0.168$^{8}$\\  
HD~197027 & 0.171$\pm$0.005 & 6 & 0.012 & \\  
\hline  
\end{tabular}  
\tablebib{(1)~\citet{b95}, (2)~\citet{Cincunegui2007yCat..34690309C}, (3)~\citet{Duncan2005yCat.3159.0D}, (4)~\citet{Gray2003AJ.126.2048G}, (5)~\citet{Hall2007AJ133.2206H}, (6)~\citet{Isaacson2010ApJ725875I}, (7)~\citet{Jenkins2011yCat..35319008J}, (8)~\citet{Melendez2012A&A..543A.29M}, (9)~\citet{Wright2004yCat.21520261W}.  
}  
\end{center}  
\end{table}  
  
The S$_{\rm{MWO}}$-values obtained for HD~101364 and HD~197027 are of the same   
order as the current activity levels of the Sun and 18~Sco. For HD~197027   
this result was completely unexpected because the activity level for a  
supposedly old solar-like star is expected   
to be the near the basal flux level \citep{schroeder2013A&A554A50S} at   
S$_{\rm{MWO}}$=0.144 \citep{mittag2013A&A549A117M}, which  
is the case for the more evolved solar-like star 51~Peg.   
However, on a longer term scale, the S$_{\rm{MWO}}$ values we sampled   
over a period of about 20 months are only  snapshots and may   
therefore not entirely represent the true stellar mean activity.   
For the Sun and 18~Sco we know that the activity in the Ca~II~H\&K lines   
show long-term periodic variations on timescales of 11 and 7.1 yr,   
respectively \citep{b95,hall2007AJ133862H,Hall2007AJ133.2206H}.    
This might in principle also be true for HD~101364 and HD~197027,  
for which we currently have only a few measurements. Consequently, some  
uncertainty remains as to how representative the obtained S$_{\rm{MWO}}$
values  
of the long-term activity level of HD~101364 and HD~197027 really are; they could,  
by bad luck, instead be indicative of a maximum or a minimum state. More  
\begin{figure}  
\centering  
\includegraphics[scale=0.35, angle=90]{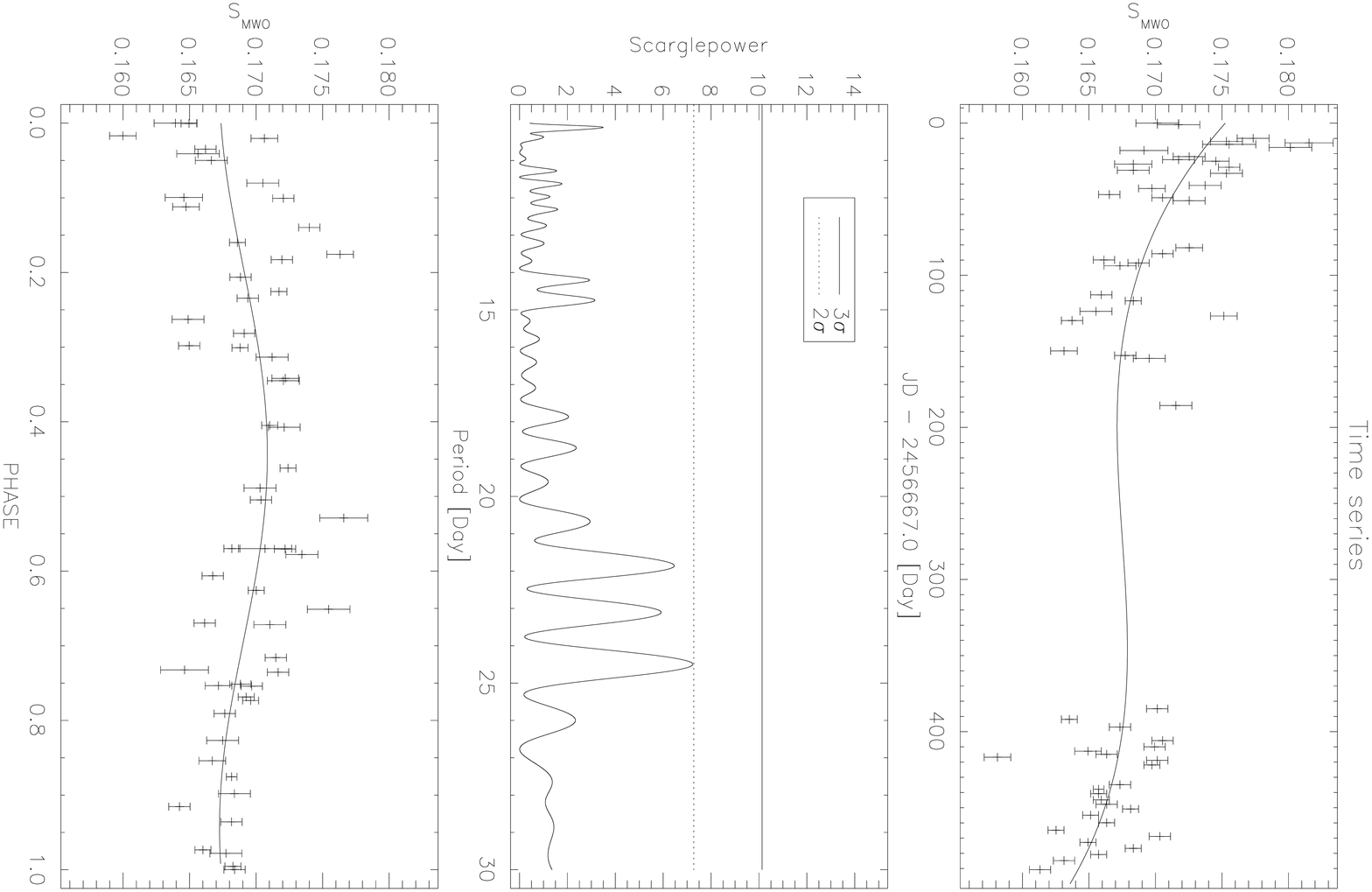}  
\caption{Upper panel: S$_{\rm{MWO}}$ time series of 18~Sco: The solid line represents   
a third-order polynomial fit, which was used for de-trending.  
Middle panel: Periodogram of the time series.  
Bottom panel: Phase-folded and de-trended time series.}  
\label{smwo_time_HD146233}  
\end{figure}  
observations of these stars are therefore required.  
  
\section{Rotation periods}  
  
According to the rotation-age-activity paradigm, the rotation period of a   
star is also a good indicator of its chromospheric activity level   
\citep{noyes1984}. Unfortunately, the rotation periods of  
HD~101364 and HD~197027 are still unknown, and our time series  
are still insufficient for a good determination.   
By contrast, the solar rotation period is of course very well determined, and   
for 18~Sco a period of 22.7 days has been established by  
\cite{Petit2008MNRAS.388...80P} with a quoted error of 0.5 days. To determine this period, \cite{Petit2008MNRAS.388...80P} used ten observations that covered   
about one and a half rotations.  
By comparison, \citet{Saar1997MNRAS.284..803S} estimated the rotation period of 18~Sco  
to be 23.7 days, using the Noyes relation \citep{noyes1984} between   
rotation period and Rossby number, respectively, and log R$^{'}_{\rm{HK}}$   
(Saar, private communication).   
For 51~Peg, a rather long period of 37.0 days was   
\begin{figure}  
\centering  
\includegraphics[scale=0.35, angle=90]{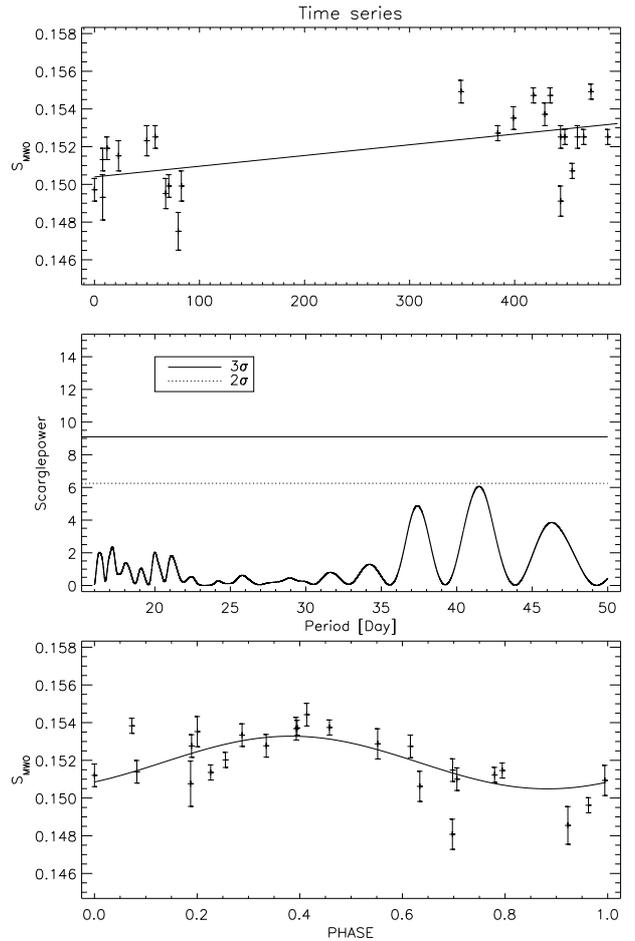}  
\caption{S$_{\rm{MWO}}$ time series of 51~Peg: The solid line represents   
the polynomial fit of the first order, which was used for the de-trending.   
In the middle the periodogram of the time series is shown, and the lower   
panel shows the phase-folded and de-trended time series.}  
\label{smwo_time_HD217014}  
\end{figure}  
found by \citet{Baliunas1996ApJ...457L..99B} from Ca~II~H\&K measurements, but   
the accuracy and significance of this period has not been discussed   
by \cite{Baliunas1996ApJ...457L..99B}.  
Furthermore, a period of 29.7 days was calculated for   
51~Peg by \cite{noyes1984}, while  \citet{Soderblom1985AJ.....90.2103S} obtained a period of $\text{about }$22 days.  
These calculated periods are shorter by 20$\%$ and 40$\%$, respectively,   
than the period measured by \citet{Baliunas1996ApJ...457L..99B}.  
  
Using the Lomb-Scargle (LS) method \citep{horne-b1986}, we analysed our   
S$_{\rm{MWO}}$ time series to check to which extent we  
 recovered the known rotation period for the Sun. We checked the 
published rotation periods for 18~Sco and 51~Peg with a TIGRE-derived   
LS periodogram.   
  
In our solar daily S-index series from 2014, we found a period of   
$\approx$26.7 days with a 3$\sigma$ significance, which is well  
consistent with the solar synodic rotation period of 27.28 days    
\citep{Cox2000asqu}.  
  
In Fig.~\ref{smwo_time_HD146233} we present the TIGRE S$_{\rm{MWO}}$ time   
series for 18~Sco. The raw data and a trend line are plotted in the  
top panel, the LS periodogram is shown in the middle panel,   
and in the bottom panel we show the time series folded with the most significant period;  
de-trending was accomplished by using    
a third-order polynomial. The periodogram reveals   
three peaks (21.9, 23.1, and 24.5 days) with a significance above $1\sigma$   
around the value of 22.7 days derived by \citet{Petit2008MNRAS.388...80P}.   
The highest peak in our   
periodogram (see Fig. \ref{smwo_time_HD146233}) results in a period of 24.5 days   
with a significance of 95.3~$\%$, that is, almost at the $2\sigma$ level.   
  
In Fig.~\ref{smwo_time_HD217014} we present our TIGRE 51~Peg S$_{\rm{MWO}}$ data  
and period analysis in the same way as in Fig.~\ref{smwo_time_HD146233}.  
Here we find a small linear trend in the time series, and the LS  
periodogram then shows two peaks above the $1\sigma$ level, pointing at   
periods of 37.4 days (formal significance 83~$\%$) and 41.5 days   
(formal significance 94.5~$\%$), both of which are comparable with   
the value published by \cite{Baliunas1996ApJ...457L..99B}.   
  
While the formal significance levels of the periods determined for 18~Sco and   
51~Peg are below the $2\sigma$ level, the determined periods compare   
well with the literature values. Therefore, we conclude that our Lomb-Scargle 
results are likely consistent with the true rotational periods of   
the observed stars, and in particular, that 51~Peg rotates significantly  
more slowly than the Sun, while 18~Sco is a somewhat faster rotator.  
  
\section{Evolutionary ages, or how good is a solar twin?}  
  
Whether a star serves as a good solar twin does not only depend   
on the physical parameters luminosity $L$, effective temperature   
$T_{\rm eff}$ , and metallicity ([Fe/H]), the star should also   
have a mass and age very similar to that of the Sun.   
On or near the MS, younger stars with slightly   
higher mass have the same effective temperature and are only slightly more   
luminous than their older siblings. Hence, a careful age analysis  
must be an important aspect in any discussion of solar twins.     
  
Estimating stellar ages is notoriously difficult.   
The derived ages depend on the method used and to exacerbate
the problem, the uncertainties tend to significantly increase with age.   
We consider 18 Sco. Published age estimates for this star range    
from 0.29 Gyr \citep{Tsantaki2013yCat..35559150T} to 5.84 Gyr   
\citep{Takeda2008yCat..21680297T}, that is, from younger than the Hyades to  
older than the Sun.~ The youngest age is in   
clear contrast to all other published values, which give 18~Sco a more   
mature evolutionary status, closer to that of the Sun; our own result   
(see below) of 5.1 Gyr confirms this.   
\begin{table*}[!t]  
\caption{Summary of the used stellar parameters and our results of the age estimations.} 
\label{tab4}  
\smallskip  
\begin{center}  
\begin{tabular}{l c c c c c c c c c}  
\hline  
\hline  
\noalign{\smallskip}  
Name &  $M_{\rm{V}}$ & [Fe/H] & $log L/L_{\odot}$ & $\overline{T}_{\rm eff}$ [K] & $\sigma_{T_{\rm eff}}$ [K]& $M_{\odot}$ & Z & Age [Gyr] & Ref Age [Gyr]  \\  
\hline  
\noalign{\smallskip}  
Sun       & 4.82$^{1}$      & 0.0        & 0               & 5777$^{1}$   &    & 1.00 & 0.0185 & 4.6         & 7.5\\  
18~Sco    & 4.77$\pm$0.01  & 0.03$^{4}$  & 0.018$\pm$0.004 & 5789$\pm$10 & 46 & 1.00 & 0.0185 & 5.1$\pm$1.1 & 7.5 \\  
51~Peg    & 4.48$\pm$0.01  & 0.2$^{4}$   & 0.135$\pm$0.005 & 5768$\pm$8  & 41 & 1.11 & 0.0296 & 6.1$\pm$0.6 & 6.1 \\  
HD~101364 & 4.65$\pm$0.09  & 0.02$^{2}$  & 0.07$\pm$0.04   & 5805$\pm$11 & 23 & 0.99 & 0.0185 & 7.1$\pm$1.5 & 8.2 \\  
HD~197027 & 4.72$\pm$0.16  &-0.013$^{3}$ & 0.04$\pm$0.07   & 5777$\pm$31 & 63 & 0.98 & 0.0185 & 7.2$\pm$3.1 & 8.5 \\  
\hline  
\end{tabular}
\tablefoot{Here, we list the calculated absolute visual magnitude, metallicity,   
calculated luminosity, mean effective temperature, the standard deviation ($\sigma$) 
of the effective temperature, mass, metallicity Z, the absolute ages and the reference ages.
The reference age is the age at the defined turn-around point of the evolutionary tracks.}
\tablebib{(1)~\cite{Cox2000asqu}, (2)~\cite{Melendez2012A&A..543A.29M},   
(3)~\cite{Monroe2013ApJ774L32M}, and (4)~\cite{Valenti2005yCat..21590141V}.   
}  
\end{center}  
\end{table*}  
  
To obtain a consistent interpretation of the HRD positions of all  
our sample stars and derive their absolute ages  
in a consistent manner, we employed our own evolutionary tracks
that were computed   
with the Cambridge Stellar Evolution Code\footnote[1]{http://www.ast.cam.ac.uk/$^\sim$stars/}   
\citep{Pols1997MNRAS.289.869P}, which is  
an updated version of the work of  \cite{Eggleton1971MNRAS.151..351E}.  
These evolutionary tracks, with appropriate choices of mass and metallicity,   
were then compared with the observed luminosities $L$   
and effective temperatures $T_{\rm eff}$ of the stars, derived here  
from photometric data and trigonometric parallaxes.     
  
Our Sun presents a way of testing this approach. An age   
determination based on isotopes of the oldest solar system meteorites leads to an empirical   
solar age of around 4.6 Gyr \citep{2010NatGe...3..637B}.  
On the other hand, the accuracy of evolutionary models still depends on   
their empirical calibration, despite the significant improvements in the opacity   
tables and the description of the equation of state, in particular for   
cooler plasmas. The main problem remaining, however, is the approximate   
description of convection, which still uses the mixing length approach,   
where the scaling factor (the mixing length) has to be calibrated   
against real stars and their physical properties. The uncertainty in such   
stellar data used for calibration leaves us with typical inconsistencies   
of the absolute age-scale of the order of 0.1 Gyr for models of   
solar-type stars.  Hence, we individually calibrated   
the evolutionary model used here for the Sun, which   
now suggests a age of around 4.6 Gyr.
  
The parallax errors are most important because they affect the
luminosity and consequently the age.   
The uncertainty in $T_{\rm{eff}}$, by contrast, translates mostly into   
a necessary variation of the stellar mass. We derived this in steps of  
$\Delta0.005$M$_{\odot}$ for 51 Peg and $\Delta0.01$M$_{\odot}$   
for the other stars to cover the respective error margin around $T_{\rm{eff}}$.   
However, the uncertainty in metallicity also   
causes a shift in $T_{\rm{eff}}$ of the respective evolutionary tracks  
and so adds to the effective uncertainty of the derived best-matching mass   
and age. Hence, the total error of the age was estimated   
by error propagation, using a whole range of such appropriate evolutionary   
tracks.  
  
In Table~\ref{tab4} we list the absolute visual magnitudes $M_{\rm{V}}$   
metallicities, a mean $T_{\rm{eff}}$ , and the estimated ages of our sample stars.   
For 18~Sco, 51~Peg, HD~101364, and   
HD~197027, the former were calculated from the revised  
HIPPARCOS parallaxes \citep{van_Leeuwen2008yCat.1311.0V},   
while photometric data (V and B-V values) were taken   
from the HIPPARCOS catalogue \citep{HIPPARCOS1997ESA}.  
  
The $T_{\rm{eff}}$ values used for 18~Sco, 51~Peg, HD~101364, and   
HD~197027 are (unweighted) averages from all entries listed   
in the Pastel catalogue (Version 2013-04-01;   
\cite{Soubiran2010yCat....102029S}),   
and the respective standard deviation was used to estimate the  
uncertainty of $T_{\rm{eff}}$. For HD~197027 we also included the value  
$T_{\rm{eff}}$ = 5723\,K derived by \cite{Monroe2013ApJ774L32M}.   
  
The luminosities $L/L_{\odot}$ were then derived directly from the resulting   
\begin{figure}  
\centering  
\includegraphics[scale=0.3]{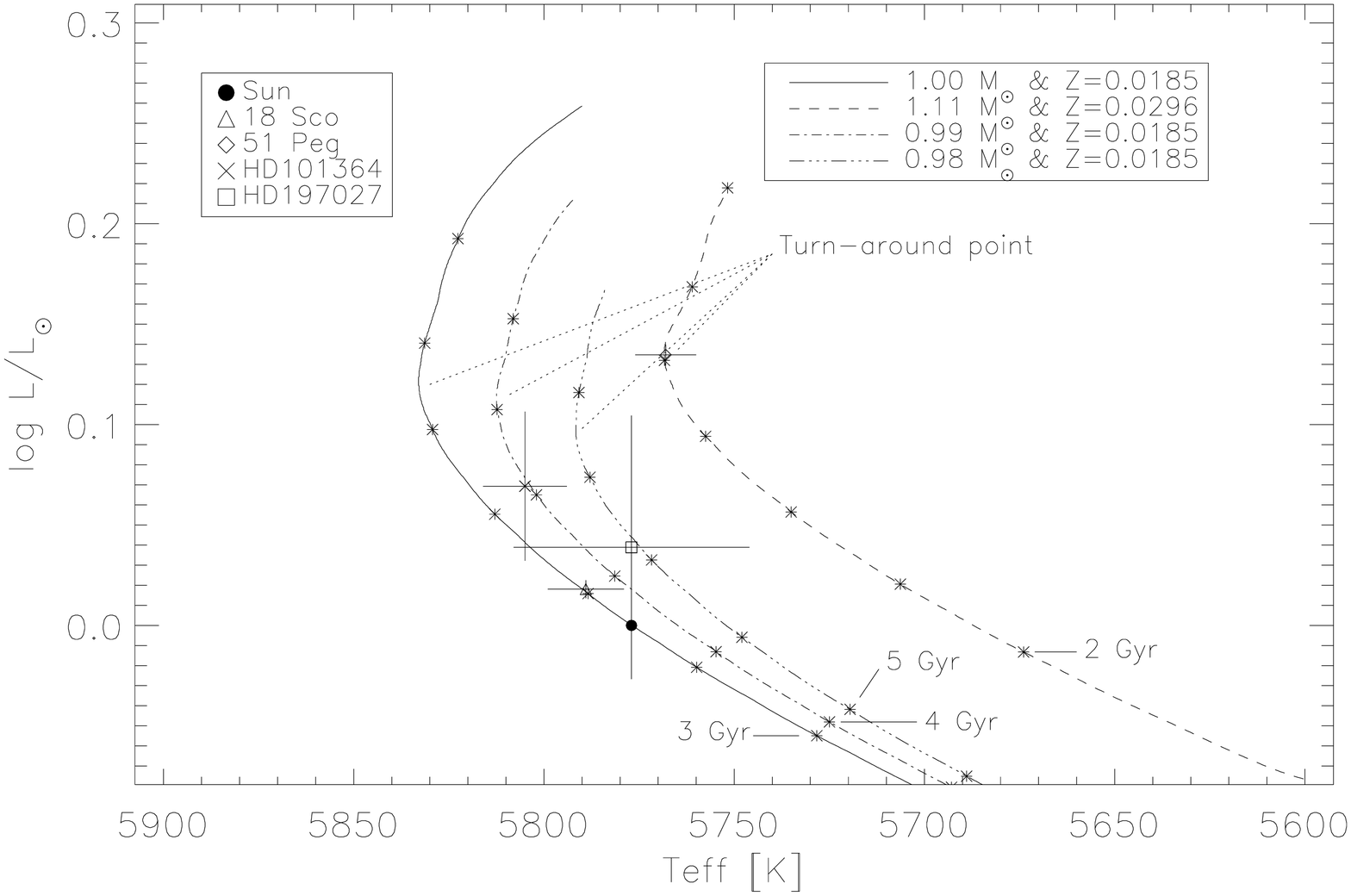}  
\caption{Luminosity vs. effective temperature of the stars.   
The solid line shows the evolutionary track for 1$M_{\odot}$,   
the dash-dotted line for 0.99$M_{\odot}$ , and the dash-triple-dotted line for   
0.98$M_{\odot}$ with Z=0.0185. The  dashed line denotes the evolutionary track for   
1.11$M_{\odot}$ and Z=0.0196. The asterisk at the evolutionary   
track marks a time step of 1 Gyr, and the first step is labelled with   
the age.}  
\label{Fig4}  
\end{figure}  
absolute visual magnitudes $M_{\rm{v}}$ by assuming that the bolometric correction   
BC is essentially the same for all sample stars. This simplification   
is justified because all $T_{\rm{eff}}$ values are very comparable within  
their uncertainties, and any real differences would cause the BCs to differ by  
only a few hundredths of a magnitude. In comparison to the propagated   
parallax error, this error is negligible.    
  
To represent the different metallicities by appropriate evolutionary   
tracks, we found that the evolutionary track with Z=0.0185   
(as for the Sun, equivalent to [Fe/H]=0.00) represents our host star and   
its twins. For the stellar mass we found that the evolutionary tracks   
with 0.99$M_{\odot}$ for HD~101364 and with 0.98$M_{\odot}$ for HD~197027   
represent the stars better than the track with 1$M_{\odot}$ like 18~Sco.  
For 51 Peg, by contrast, we used an evolutionary track computed for   
Z=0.0296 and the  
respective opacity table (equivalent to [Fe/H] $\approx$ +0.20)   
to model the effect of the much higher metallicity of this star.   
In Fig. \ref{Fig4} all sample stars   
are plotted in the HR diagram together with the evolutionary   
tracks as detailed above.  
With an estimated age of (7.1$\pm$1.5) Gyr and (7.2$\pm$3.1) Gyr,   
HD~101364 and HD~197027 seem to be significantly older than the Sun,  
while 18~Sco appears to be a much closer match to the Sun with an estimated 
age of (5.1$\pm$1.1) Gyr. However, if the age uncertainties of HD~197027 and 18~Sco 
are taken into account, they are then still comparable   
to the Sun. In particular, the age uncertainty for HD~197027 is quite large.   
To obtain a more precise age for this star, a more accurate parallax is   
needed in the first place, followed by a better $T_{\rm{eff}}$
value.   
  
In the special case of 51~Peg, we found that its higher luminosity and  
higher metallicity can be matched very well by a track of about   
1.11 solar masses with Z=0.0296. For the age of 51 Peg   
we then obtained (6.1$\pm$0.6) Gyr. The higher metallicity   
of 51~Peg makes it slightly redder than the Sun, overcompensating for   
the effect of its higher mass. Another consequence of this higher   
mass is that the   
evolutionary MS timescale of 51~Peg is shorter by 20\% than the Sun and   
the twins we discussed here.  
  
With this lower MS life-expectancy, this star is the most advanced   
on the MS when compared to the Sun and its twins, even though its   
age as such does not immediately suggest this. Hence, the very low activity  
level of 51 Peg, when compared to the Sun and its twins, suggests  
a relation between activity level and evolutionary (or relative) age   
(absolute age/reference age) and not the absolute age, when different   
stellar life expectations are compared. But how can such different evolutionary models 
be compared when for such low masses a well-defined turn-off point as a physical and visible  
(in the HRD) reference point to the MS life-expectancy is lacking?    
In these stars that are close to one solar mass, hydrogen shell-burning sets in very 
gradually, and this process leaves no suitable mark on the HRD.  
\begin{figure}  
\centering  
\includegraphics[scale=0.5]{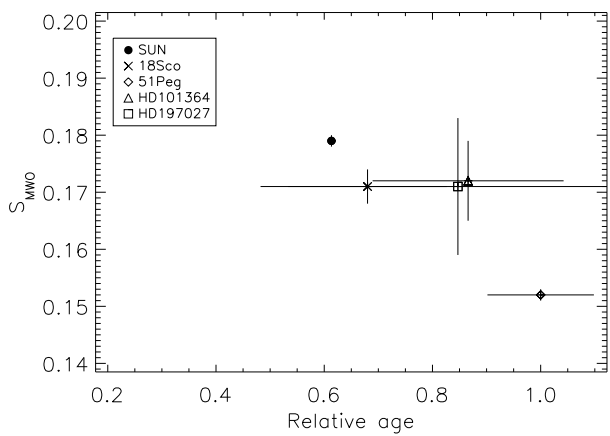}  
\caption{S$_{\rm{MWO}}$ value vs relative age. Our estimated ages are derived   
from the reference ages, see Table \ref{tab4}. For the S values the mean literature values are   
used for the Sun, 18~Sco, and 51~Peg. For HD~101364, the literature S$_{\rm{MWO}}$ value   
and our value are averaged, and only our S$_{\rm{MWO}}$-value was used for HD~197027.}  
\label{Fig5}  
\end{figure}  
Therefore, as a well-defined (albeit not physically meaningful) reference point,   
we prefer to use the turn-around point of the evolutionary tracks   
in the HRD (see label in Fig. \ref{Fig4}), which all stars of about one 
solar mass describe in the HRD during their later MS evolution.   
We use the age of the turn-around point as a reference age, although  
this location does not indicate any significant structural change in the star. However,  
it provides a well-defined location on the HRD, where the evolutionary tracks   
reverse their temperature trend with (growing) luminosity and become faster.   
For the tracks with Z=0.0185 and a stellar mass of 1M$_{\odot}$, 0.99M$_{\odot}$   
and 0.98M$_{\odot}$, the ages at these points are $\approx$7.5 Gyr, $\approx$8.2 Gyr,   
and $\approx$8.5 Gyr, while for 1.11M$_{\odot}$ and Z=0.0296, this age is   
$\approx$6.1 Gyr. In these terms, 51 Peg is much more evolved than the Sun, since  
it has reached 100\% of its turn-around age compared to the Sun
and its twins,
which are at only 60\% to 85\%    
of this age. This relative age agrees much better with the   
respective activity levels.   

This is depicted in Fig. \ref{Fig5}.
For this comparison we used our estimated relative age and 
a mean of the available literature S$_{\rm{MWO}}$ values  
for the Sun, 18~Sco, and 51~Peg, which are more representative   
for their activity in the long term than our current snapshot S$_{\rm{MWO}}$ values.   
For HD~197027, however, by lack of other choice we used our   
own S$_{\rm{MWO}}$ value, while for HD~101364, we averaged the mean   
literature S$_{\rm{MWO}}$ value obtained from \citet{Melendez2012A&A..543A.29M} with   
our measurements.   
  
\section{Discussion and conclusion}  
  
Chromospheric activity is a phenomenon showing variability  
on long timescales (years and to centuries) caused by activity cycles   
and short timescales (days to years) produced by the evolution and   
rotational modulation of individual active regions. Hence,   
snapshot observations such as we used in this study may under- or overestimate the   
long-term level of chromospheric activity even though we averaged over   
many different nights. For more certainty, long-term monitoring is therefore  
necessary.  As an example, we note that the   
long-term S-average from 1966-1992 \citep{b95} for the Sun is close to 0.18, covering   
three solar   
activity maxima {\it and} minima. Around the current relatively weak   
maximum, we record by contrast an average solar S$_{\rm{MWO}}$
value of only   
about 0.17. We can rule out that this significant difference is caused   
by our calibration procedure because we used the same stars for our transformation  
to Mt. Wilson S$_{\rm{MWO}}$ values as were used by O.C. Wilson and his   
group.   
  
In particular, for modestly active stars with S$<$0.2, the standard deviation   
of the residuals of our calibrated S$_{\rm{MWO}}$ values compared
with those from   
Baliunas et al. is as small as 0.004, and the mean variation amounts to only 2.3$\%$.   
Furthermore, during the period when we measured the solar S$_{\rm{MWO}}$
value of 0.17,   
sunspot counts were also on a lower level than in the solar   
cycles of 1966-1992. We therefore conclude that this difference is real  
and represents a real change of the solar activity on a longer (decades to   
century) timescale.   
  
This casts some doubt as to how well the present snapshot  
S$_{\rm{MWO}}$ value of 0.17 represents the long-term solar activity.   
However, when modern sunspot records are compared with historic observations,   
all uncertainties included,  
(SIDC 2015, communication of July 1 \footnote{http://sidc.oma.be/press/01/welcome.html})  
the second half of the past century (when \citet{b95} obtained their  
S$_{\rm{MWO}}$ value of 0.18 )  
appears to have been above-average and so may portray a slightly too young Sun.   
  
Interestingly, the snapshot observations   
of HD~101364 and HD~197027 we presented here show that their activity levels are   
more or less comparable with that of the current Sun and that of 18~Sco,   
but all of them clearly differ from 51~Peg, which exhibits a   
significantly and consistently lower activity.  
Exclusively from the point of view of stellar activity levels,   
we may therefore conclude that HD~101364 and HD~197027 are younger in the context   
of stellar evolution than the more evolved solar-like star 51~Peg, which has already 
come quite close to the basal flux level of its S-index.   
Compared to the Sun and 18~Sco, however,   
it might be assumed that HD~101364 and HD~197027 have a comparable   
age or are slightly older.   
  
Using stellar ages obtained from the comparison of matching evolutionary tracks  
with the observed stellar HRD positions, we find that 51~Peg is,   
in absolute terms, not the oldest object in this small sample.   
However, when the faster MS evolution of 51~Peg is considered,    
we can show the same trend in the evolutionary advance as the one shown by   
the observed activity levels (see Fig. \ref{Fig5}) because its
mass is higher than those of the other four stars. These results    
confirm earlier work, that is,  
a better correspondence between activity level and {\it relative} MS age instead of absolute age, as shown by   
\citet{Reiners2012ApJ...746...43R} and \citet{schroeder2013A&A554A50S}.  
  
Another important characteristic of a star is its rotation period.   
The rotation period is correlated with the chromospheric activity and with stellar age. 
This can also be seen in our very small sample of only three   
periods, which is in sufficient to show this correlation   
clearly. Nevertheless, we may say that the Sun and 18~Sco are comparable   
in age, in chromospheric activity, and in their rotation periods.   
Furthermore, the rotation of the Sun and 18~Sco is clearly   
faster than that of the evolved star 51~Peg. This finding agrees with   
the relation between evolutionary (relative) age and activity level shown   
in  Fig.~\ref{Fig5}.  
  
It is obvious from Fig. \ref{Fig5} that the same relation   
holds for the other two solar twins that we discussed here, HD~101364 and HD~197027.  
Even though this must still be confirmed by extending this study to a   
larger sample of solar-like stars, we may already conclude that first, the  
chromospheric activity level is a good age indicator, even for   
advanced evolutionary states, and that second, the two new   
solar-twin candidates (HD~101364 and HD~197027) and 18~Sco we discussed
here are suitable   
matches to our Sun with respect to mass, metallicity, activity, and   
evolutionary age, even if they are perhaps slightly older, and  
third, in the search for a perfect twin, stellar age should be   
given more consideration    
because otherwise the activity level would not match that of the Sun.

\begin{acknowledgements}  
This research has made valuable use of the VizieR catalogue access tool,   
CDS, Strasbourg, France. The original description of the VizieR service   
was published in A\&AS 143, 23.  We are grateful for the   
financial support received from the  
DFG project No. SCHM 1032/49-1, and we further acknowledge travel support   
by bilateral grants CONACyT-DFG (No.192334) and CONACyT-DAAD (No. 207772)   
in the past two years, as well as by the CONACyT mobility grant (No. 207662).  
\end{acknowledgements}


\begin{thebibliography}{38}
\expandafter\ifx\csname natexlab\endcsname\relax\def\natexlab#1{#1}\fi

\bibitem[{{Baliunas} {et~al.}(1996){Baliunas}, {Sokoloff}, \&
  {Soon}}]{Baliunas1996ApJ...457L..99B}
{Baliunas}, S., {Sokoloff}, D., \& {Soon}, W. 1996, \apjl, 457, L99

\bibitem[{{Baliunas} {et~al.}(1995){Baliunas}, {Donahue}, {Soon}, {Horne},
  {Frazer}, {Woodard-Eklund}, {Bradford}, {Rao}, {Wilson}, {Zhang}, {Bennett},
  {Briggs}, {Carroll}, {Duncan}, {Figueroa}, {Lanning}, {Misch}, {Mueller},
  {Noyes}, {Poppe}, {Porter}, {Robinson}, {Russell}, {Shelton}, {Soyumer},
  {Vaughan}, \& {Whitney}}]{b95}
{Baliunas}, S.~L., {Donahue}, R.~A., {Soon}, W.~H., {et~al.} 1995, \apj, 438,
  269

\bibitem[{{Bouvier} \& {Wadhwa}(2010)}]{2010NatGe...3..637B}
{Bouvier}, A. \& {Wadhwa}, M. 2010, Nature Geoscience, 3, 637

\bibitem[{{Cincunegui} {et~al.}(2007){Cincunegui}, {Diaz}, \&
  {Mauas}}]{Cincunegui2007yCat..34690309C}
{Cincunegui}, C., {Diaz}, R.~F., \& {Mauas}, P.~J.~D. 2007, VizieR Online Data
  Catalog, 346, 90309

\bibitem[{{Cox}(2000)}]{Cox2000asqu}
{Cox}, A.~N. 2000, {Allen's astrophysical quantities}

\bibitem[{{Duncan} {et~al.}(1991){Duncan}, {Vaughan}, {Wilson}, {Preston},
  {Frazer}, {Lanning}, {Misch}, {Mueller}, {Soyumer}, {Woodard}, {Baliunas},
  {Noyes}, {Hartmann}, {Porter}, {Zwaan}, {Middelkoop}, {Rutten}, \&
  {Mihalas}}]{duncan1991}
{Duncan}, D.~K., {Vaughan}, A.~H., {Wilson}, O.~C., {et~al.} 1991, \apjs, 76,
  383

\bibitem[{{Duncan} {et~al.}(2005){Duncan}, {Vaughan}, {Wilson}, {Preston},
  {Frazer}, {Lanning}, {Misch}, {Mueller}, {Soyumer}, {Woodard}, {Baliunas},
  {Noyes}, {Hartmann}, {Porter}, {Zwaan}, {Middelkoop}, {Rutter}, \&
  {Mihalas}}]{Duncan2005yCat.3159.0D}
{Duncan}, D.~K., {Vaughan}, A.~H., {Wilson}, O.~C., {et~al.} 2005, VizieR
  Online Data Catalog, 3159, 0

\bibitem[{{Eggleton}(1971)}]{Eggleton1971MNRAS.151..351E}
{Eggleton}, P.~P. 1971, \mnras, 151, 351

\bibitem[{{ESA}(1997)}]{HIPPARCOS1997ESA}
{ESA}, ed. 1997, ESA Special Publication, Vol. 1200, {The HIPPARCOS and TYCHO
  catalogues. Astrometric and photometric star catalogues derived from the ESA
  HIPPARCOS Space Astrometry Mission}

\bibitem[{{Gray} {et~al.}(2003){Gray}, {Corbally}, {Garrison}, {McFadden}, \&
  {Robinson}}]{Gray2003AJ.126.2048G}
{Gray}, R.~O., {Corbally}, C.~J., {Garrison}, R.~F., {McFadden}, M.~T., \&
  {Robinson}, P.~E. 2003, \aj, 126, 2048

\bibitem[{{Hall} {et~al.}(2007{\natexlab{a}}){Hall}, {Henry}, \&
  {Lockwood}}]{Hall2007AJ133.2206H}
{Hall}, J.~C., {Henry}, G.~W., \& {Lockwood}, G.~W. 2007{\natexlab{a}}, \aj,
  133, 2206

\bibitem[{{Hall} {et~al.}(2007{\natexlab{b}}){Hall}, {Lockwood}, \&
  {Skiff}}]{hall2007AJ133862H}
{Hall}, J.~C., {Lockwood}, G.~W., \& {Skiff}, B.~A. 2007{\natexlab{b}}, \aj,
  133, 862

\bibitem[{Hauschildt {et~al.}(1999)Hauschildt, Allard, \&
  Baron}]{hauschildt1999}
Hauschildt, P.~H., Allard, F., \& Baron, E. 1999, ApJ, 512, 377

\bibitem[{{Horne} \& {Baliunas}(1986)}]{horne-b1986}
{Horne}, J.~H. \& {Baliunas}, S.~L. 1986, \apj, 302, 757

\bibitem[{{Isaacson} \& {Fischer}(2010)}]{Isaacson2010ApJ725875I}
{Isaacson}, H. \& {Fischer}, D. 2010, \apj, 725, 875

\bibitem[{{Jenkins} {et~al.}(2011){Jenkins}, {Murgas}, {Rojo}, {Jones},
  {Day-Jones}, {Jones}, {Clarke}, {Ruiz}, \&
  {Pinfield}}]{Jenkins2011yCat..35319008J}
{Jenkins}, J.~S., {Murgas}, F., {Rojo}, P., {et~al.} 2011, VizieR Online Data
  Catalog, 353, 19008

\bibitem[{{Mamajek} \& {Hillenbrand}(2008)}]{Mamajek2008ApJ687.1264M}
{Mamajek}, E.~E. \& {Hillenbrand}, L.~A. 2008, \apj, 687, 1264

\bibitem[{{Melendez} {et~al.}(2012){Melendez}, {Bergemann}, {Cohen}, {Endl},
  {Karakas}, {Ram{\'{\i}}rez}, {Cochran}, {Yong}, {MacQueen}, {Kobayashi}, \&
  {Asplund}}]{Melendez2012A&A..543A.29M}
{Melendez}, J., {Bergemann}, M., {Cohen}, J.~G., {et~al.} 2012, \aap, 543, A29

\bibitem[{{Mittag} {et~al.}(2013){Mittag}, {Schmitt}, \&
  {Schr{\"o}der}}]{mittag2013A&A549A117M}
{Mittag}, M., {Schmitt}, J.~H.~M.~M., \& {Schr{\"o}der}, K.-P. 2013, \aap, 549,
  A117

\bibitem[{{Monroe} {et~al.}(2013){Monroe}, {Mel{\'e}ndez}, {Ram{\'{\i}}rez},
  {Yong}, {Bergemann}, {Asplund}, {Bedell}, {Tucci Maia}, {Bean}, {Lind},
  {Alves-Brito}, {Casagrande}, {Castro}, {do Nascimento}, {Bazot}, \&
  {Freitas}}]{Monroe2013ApJ774L32M}
{Monroe}, T.~R., {Mel{\'e}ndez}, J., {Ram{\'{\i}}rez}, I., {et~al.} 2013,
  \apjl, 774, L32

\bibitem[{{Noyes} {et~al.}(1984){Noyes}, {Hartmann}, {Baliunas}, {Duncan}, \&
  {Vaughan}}]{noyes1984}
{Noyes}, R.~W., {Hartmann}, L.~W., {Baliunas}, S.~L., {Duncan}, D.~K., \&
  {Vaughan}, A.~H. 1984, \apj, 279, 763

\bibitem[{{Petit} {et~al.}(2008){Petit}, {Dintrans}, {Solanki}, {Donati},
  {Auri{\`e}re}, {Ligni{\`e}res}, {Morin}, {Paletou}, {Ramirez Velez},
  {Catala}, \& {Fares}}]{Petit2008MNRAS.388...80P}
{Petit}, P., {Dintrans}, B., {Solanki}, S.~K., {et~al.} 2008, \mnras, 388, 80

\bibitem[{{Piskunov} \& {Valenti}(2002)}]{REDUCE2002A&A385.1095P}
{Piskunov}, N.~E. \& {Valenti}, J.~A. 2002, \aap, 385, 1095

\bibitem[{{Pols} {et~al.}(1997){Pols}, {Tout}, {Schroder}, {Eggleton}, \&
  {Manners}}]{Pols1997MNRAS.289.869P}
{Pols}, O.~R., {Tout}, C.~A., {Schroder}, K.-P., {Eggleton}, P.~P., \&
  {Manners}, J. 1997, \mnras, 289, 869

\bibitem[{{Porto de Mello} \& {da
  Silva}(1997)}]{Porto_de_Mello1997ApJ...482L..89P}
{Porto de Mello}, G.~F. \& {da Silva}, L. 1997, \apjl, 482, L89

\bibitem[{{Ram{\'{\i}}rez} {et~al.}(2014){Ram{\'{\i}}rez}, {Mel{\'e}ndez},
  {Bean}, {Asplund}, {Bedell}, {Monroe}, {Casagrande}, {Schirbel}, {Dreizler},
  {Teske}, {Tucci Maia}, {Alves-Brito}, \& {Baumann}}]{Ramirez2014A&A572A48R}
{Ram{\'{\i}}rez}, I., {Mel{\'e}ndez}, J., {Bean}, J., {et~al.} 2014, \aap, 572,
  A48

\bibitem[{{Reiners} \& {Mohanty}(2012)}]{Reiners2012ApJ...746...43R}
{Reiners}, A. \& {Mohanty}, S. 2012, \apj, 746, 43

\bibitem[{{Saar} \& {Osten}(1997)}]{Saar1997MNRAS.284..803S}
{Saar}, S.~H. \& {Osten}, R.~A. 1997, \mnras, 284, 803

\bibitem[{{Schmitt} {et~al.}(2014){Schmitt}, {Schr{\"o}der}, {Rauw},
  {Hempelmann}, {Mittag}, {Gonz{\'a}lez-P{\'e}rez}, {Czesla}, {Wolter}, {Jack},
  {Eenens}, \& {Trinidad}}]{Schmitt2014AN335787S}
{Schmitt}, J.~H.~M.~M., {Schr{\"o}der}, K.-P., {Rauw}, G., {et~al.} 2014,
  Astronomische Nachrichten, 335, 787

\bibitem[{{Schr{\"o}der} {et~al.}(2013){Schr{\"o}der}, {Mittag}, {Hempelmann},
  {Gonz{\'a}lez-P{\'e}rez}, \& {Schmitt}}]{schroeder2013A&A554A50S}
{Schr{\"o}der}, K.-P., {Mittag}, M., {Hempelmann}, A.,
  {Gonz{\'a}lez-P{\'e}rez}, J.~N., \& {Schmitt}, J.~H.~M.~M. 2013, \aap, 554,
  A50

\bibitem[{{Soderblom}(1985)}]{Soderblom1985AJ.....90.2103S}
{Soderblom}, D.~R. 1985, \aj, 90, 2103

\bibitem[{{Soubiran} {et~al.}(2010){Soubiran}, {Le Campion}, {Cayrel de
  Strobel}, \& {Caillo}}]{Soubiran2010yCat....102029S}
{Soubiran}, C., {Le Campion}, J.-F., {Cayrel de Strobel}, G., \& {Caillo}, A.
  2010, VizieR Online Data Catalog, 1, 2029

\bibitem[{{Takeda} {et~al.}(2008){Takeda}, {Ford}, {Sills}, {Rasio}, {Fischer},
  \& {Valenti}}]{Takeda2008yCat..21680297T}
{Takeda}, G., {Ford}, E.~B., {Sills}, A., {et~al.} 2008, VizieR Online Data
  Catalog, 216, 80297

\bibitem[{{Tsantaki} {et~al.}(2013){Tsantaki}, {Sousa}, {Adibekyan}, {Santos},
  {Mortier}, \& {Israelian}}]{Tsantaki2013yCat..35559150T}
{Tsantaki}, M., {Sousa}, S.~G., {Adibekyan}, V.~Z., {et~al.} 2013, VizieR
  Online Data Catalog, 355, 59150

\bibitem[{{Valenti} \& {Fischer}(2005)}]{Valenti2005yCat..21590141V}
{Valenti}, J.~A. \& {Fischer}, D.~A. 2005, VizieR Online Data Catalog, 215,
  90141

\bibitem[{{van Leeuwen}(2008)}]{van_Leeuwen2008yCat.1311.0V}
{van Leeuwen}, F. 2008, VizieR Online Data Catalog, 1311, 0

\bibitem[{{Vaughan} {et~al.}(1978){Vaughan}, {Preston}, \&
  {Wilson}}]{Vaughan1978PASP90267V}
{Vaughan}, A.~H., {Preston}, G.~W., \& {Wilson}, O.~C. 1978, \pasp, 90, 267

\bibitem[{{Wright} {et~al.}(2004){Wright}, {Marcy}, {Butler}, \&
  {Vogt}}]{Wright2004yCat.21520261W}
{Wright}, J.~T., {Marcy}, G.~W., {Butler}, R.~P., \& {Vogt}, S.~S. 2004, VizieR
  Online Data Catalog, 215, 20261

\end{thebibliography}

\end{document}